\documentclass[manuscript]{acmart}

\usepackage{caption}
\usepackage{subcaption}
\usepackage{graphicx}

\AtBeginDocument{%
  \providecommand\BibTeX{{%
    \normalfont B\kern-0.5em{\scshape i\kern-0.25em b}\kern-0.8em\TeX}}}

\setcopyright{acmcopyright}
\copyrightyear{2018}
\acmYear{2018}
\acmDOI{10.1145/1122445.1122456}

\acmConference[Woodstock '18]{Woodstock '18: ACM Symposium on Neural
  Gaze Detection}{June 03--05, 2018}{Woodstock, NY}
\acmBooktitle{Woodstock '18: ACM Symposium on Neural Gaze Detection,
  June 03--05, 2018, Woodstock, NY}
\acmPrice{15.00}
\acmISBN{978-1-4503-XXXX-X/18/06}

\begin{document}

%%
%% The "title" command has an optional parameter,
%% allowing the author to define a "short title" to be used in page headers.
%\title{Exploring Interface Design for Intelligent Tutoring System to Improve Student Engagement}
%\title{Improving AI Powered User Interface Design for Intelligent Tutoring System Enhances Student Engagement}
%\title{Improving AI Powered User Interface Design for Intelligent Tutoring System Promotes Student Engagement}
%\title{Empirically Verifying Impacts of User Interface Design for Intelligent Tutoring System on Student Engagement}
%\title{Improving Diagnostic Interface for Intelligent Tutoring System Promotes Student Engagement}

\title{AI-Driven Interface Design for Intelligent Tutoring System Improves Student Engagement}

%%
%% The "author" command and its associated commands are used to define
%% the authors and their affiliations.
%% Of note is the shared affiliation of the first two authors, and the
%% "authornote" and "authornotemark" commands
%% used to denote shared contribution to the research.
\author{Byungsoo Kim}
\email{byungsoo.kim@riiid.co}
\affiliation{%
  \institution{Riiid! AI Research}
}
\author{Hongseok Suh}
\email{hongseok.suh@riiid.co}
\affiliation{%
  \institution{Riiid! AI Research}
}
\author{Jaewe Heo}
\email{jwheo@riiid.co}
\affiliation{%
  \institution{Riiid! AI Research}
}
\author{Youngduck Choi}
\email{youngduck.choi@riiid.co}
\affiliation{%
  \institution{Riiid! AI Research,}
  \institution{Yale University}
}

%%
%% By default, the full list of authors will be used in the page
%% headers. Often, this list is too long, and will overlap
%% other information printed in the page headers. This command allows
%% the author to define a more concise list
%% of authors' names for this purpose.
\renewcommand{\shortauthors}{Byungsoo, et al.}

%%
%% The abstract is a short summary of the work to be presented in the
%% article.
\begin{abstract}
An Intelligent Tutoring System (ITS) has been shown to improve students' learning outcomes by providing a personalized curriculum that addresses individual needs of every student.
However, despite the effectiveness and efficiency that ITS brings to students' learning process, most of the studies in ITS research have conducted less effort to design the interface of ITS that promotes students' interest in learning, motivation and engagement by making better use of AI features.
In this paper, we explore AI-driven design for the interface of ITS describing diagnostic feedback for students' problem-solving process and investigate its impacts on their engagement.
We propose several interface designs powered by different AI components and empirically evaluate their impacts on student engagement through \emph{Santa}, an active mobile ITS.
Controlled A/B tests conducted on more than 20K students in the wild show that AI-driven interface design improves the factors of engagement by up to 25.13\%.
\end{abstract}

%%
%% The code below is generated by the tool at http://dl.acm.org/ccs.cfm.
%% Please copy and paste the code instead of the example below.
%%

%%
%% Keywords. The author(s) should pick words that accurately describe
%% the work being presented. Separate the keywords with commas.

%%
%% This command processes the author and affiliation and title
%% information and builds the first part of the formatted document.
\maketitle

\section{Introduction}
The recent COVID-19 pandemic has caused unprecedented impact all across the globe.
With social distancing measures in place, many organizations have implemented virtual and remote services to prevent widespread infection of the disease and support the social needs of the public.
Educational systems are no exception and have changed dramatically with the distinctive rise of online learning.
Also, the demands for evaluation methods of learning outcomes that are safe, reliable and acceptable have led the educational environment to take a paradigm shift to formative assessment.

An Intelligent Tutoring System (ITS), which provides pedagogical services in an automated manner, is a promising technique to overcome the challenges post COVID-19 educational environment has brought to us.
However, despite the development and growing popularity of ITS, most of the studies in ITS research have mainly focused on diagnosing students' knowledge state and suggesting proper learning items, and less effort has been conducted to design the interface of ITS that promotes students' interest in learning, motivation and engagement by making better use of AI features.
For example, Knowledge Tracing (KT), a task of modeling students' knowledge through their learning activities over time, is a long-standing problem in the field of Artificial Intelligence in Education (AIEd).
From Bayesian Knowledge Tracing \cite{corbett1994knowledge,yudelson2013individualized} to Collaborative Filtering \cite{thai2010recommender,lee2016machine} and Deep Learning \cite{piech2015deep,zhang2017dynamic,choi2020towards,ghosh2020context}, various approaches have been proposed and KT is still being actively studied.
Learning path construction is also an essential task that ITS performs, where learning items are suggested to maximize students' learning objectives.
This task is commonly formulated as a reinforcement learning framework \cite{liu2019exploiting,huang2019exploring,bassen2020reinforcement,zhou2020improving} and is also an active research area in AIEd.
On the other hand, little works have been done in the context of the user interface for ITS including intelligent authoring shell \cite{granic2000user}, affective interface \cite{lin2014usability,lin2014influence} and usability testing \cite{chughtai2015usability,koscianski2014design,roscoe2014writing}.
Although they cover important aspects of ITS, the methods are outdated and their effectiveness is not reliable since the experiments were conducted on a small scale.

Interface of ITS not fully supportive of making AI’s analysis transparent to students adversely affects their engagement.
Accordingly, improving the interface of ITS is also closely related to explainable AI.
Explaining what exactly makes AI models arrive at their predictions and making them transparent to users is an important issue \cite{gunning2017explainable,gunning2019darpa,dove2020monsters}, and have been actively studied in both human-computer interaction \cite{abdul2018trends,kizilcec2016much,stumpf2009interacting,wang2019designing} and machine learning \cite{samek2017explainable} communities.
There are lots of works about the issue of explainability in many subfields of AI including computer vision \cite{norcliffe2018learning,fong2017interpretable,kim2017interpretable,zhang2018interpretable}, natural language processing \cite{lei2017interpretable,fyshe2015compositional,jiang2018interpretable,panigrahi2019word2sense} and speech processing \cite{ravanelli2018interpretable,korzekwa2019interpretable,sun2020fully,tan2015improving}.
Explainability in AIEd is mainly studied in the direction of giving feedback that helps students identify their strengths and weaknesses.
A method of combining item response theory with deep learning has been proposed, from which students' proficiency levels on specific knowledge concepts can be found \cite{cheng2019dirt,wang2019neural,yeung2019deep}.
Also, \cite{barria2019explaining,choi2020choose} attempted to give students insight why the system recommends a specific learning material.

In this paper, we explore AI-driven design for the interface of ITS describing diagnostic feedback for students' problem-solving process and investigate its impacts on their engagement.
We propose several interface designs composed of different AI-powered components.
Each page design couples the interface with AI features in different levels, providing different levels of information and explainability.
We empirically evaluate the impacts of each design on student engagement through \emph{Santa}, an active mobile ITS.
We consider conversion rate, Average Revenue Per User (ARPU), total profit, and the average number of free questions a student consumed as factors measuring the degree of engagement.
Controlled A/B tests conducted on more than 20K students in the wild show that AI-driven interface design improves the factors of engagement by up to 25.13\%.

\section{Santa: A Self-Study Solution Equipped with an AI Tutor}
\begin{figure*}
  \centering
  \includegraphics[width=1\columnwidth]{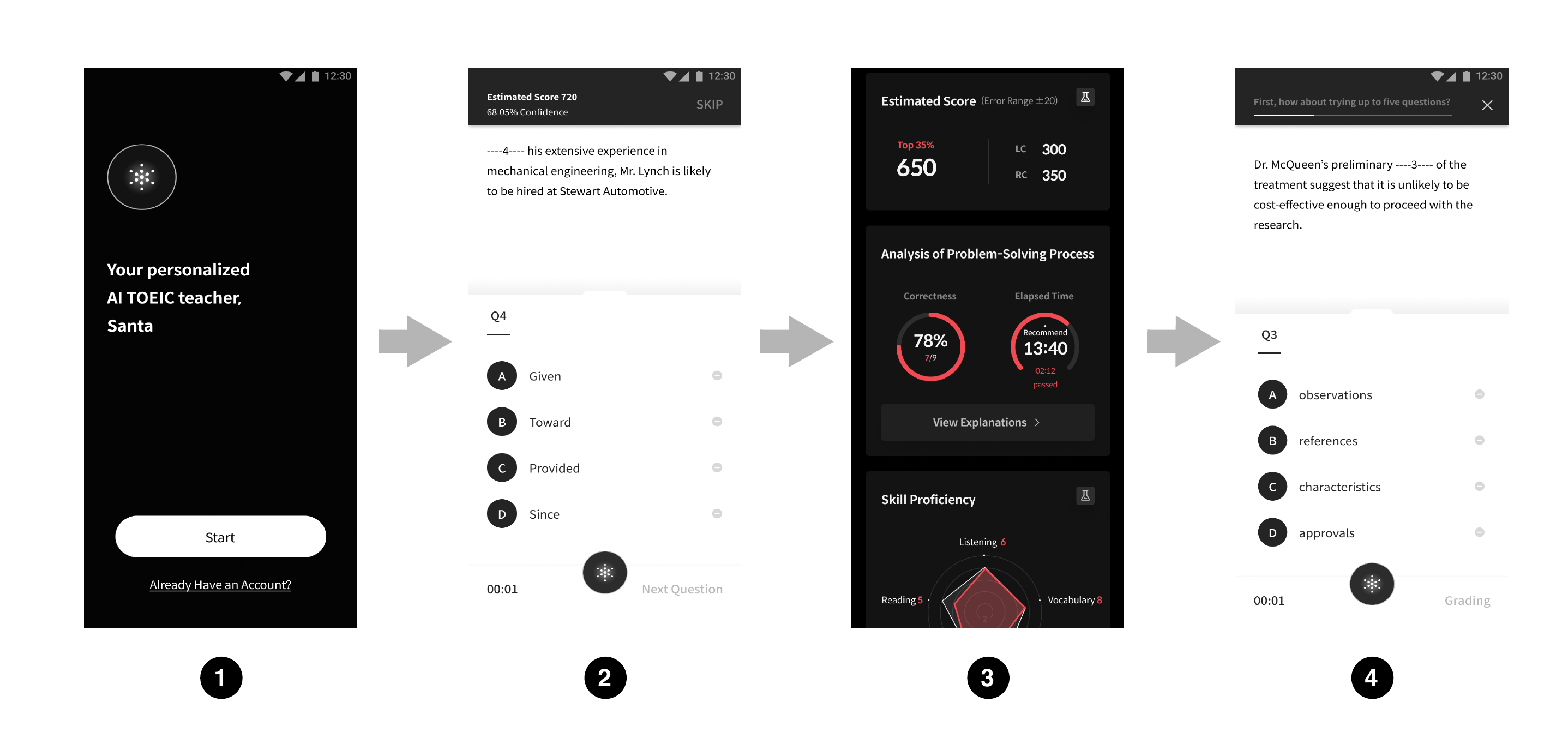}
  \caption{The flow of a user entering and interacting with \emph{Santa}}
  \label{fig:santa_flow}
\end{figure*}

\emph{Santa}\footnote{\url{https://aitutorsanta.com}} is a multi-platform AI tutoring service with more than a million users in South Korea available through Android, iOS and Web that exclusively focuses on the Test of English for International Communication (TOEIC) standardized examination.
The test consists of two timed sections named Listening Comprehension (LC) and Reading Comprehension (RC) with a total of 100 questions, and 4 or 3 parts respectively.
The final test score ranges from 10 to 990 in steps of 5 points.
Santa helps users prepare the test by diagnosing their current state and dynamically suggesting learning items appropriate for their condition.
Once a user solves each question, \emph{Santa} provides educational feedback to their responses including explanation, lecture or another question.
The flow of a user entering and using the service is described in Figure \ref{fig:santa_flow}.
When a new user first opens \emph{Santa}, they are greeted by a diagnostic test (1).
The diagnostic test consists of seven to eleven questions resembling the questions that appear on the TOEIC exam (2).
As the user progresses through the diagnostic test, \emph{Santa} records the user's activity and feeds it to a back-end AI engine that models the individual user.
At the end of the diagnostic test, the user is presented with a diagnostic page detailing the analytics of the user’s problem-solving process in the diagnostic test (3).
After that, the user may choose to continue their study by solving practice questions (4).
\section{Improving User Engagement Through an AI-driven Design for Diagnostic Interface}
\begin{figure*}
  \centering
  \begin{subfigure}[ht]{0.49\columnwidth}
  \centering
  \includegraphics[width=0.5\columnwidth]{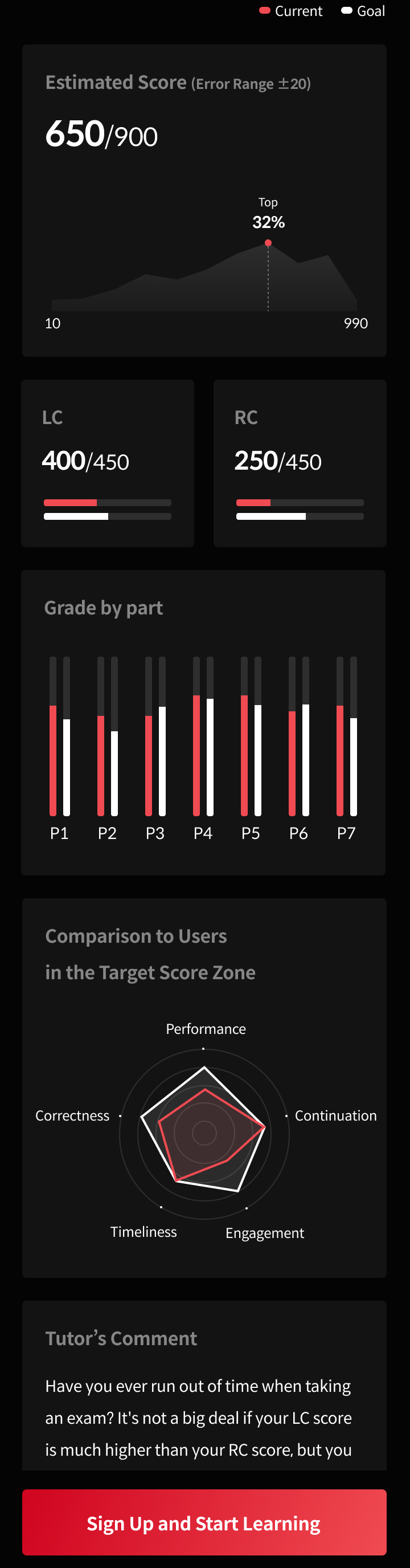}
  \caption{}
  \label{fig:page_A}
  \end{subfigure}
  \begin{subfigure}[ht]{0.49\columnwidth}
  \centering
  \includegraphics[width=1\columnwidth]{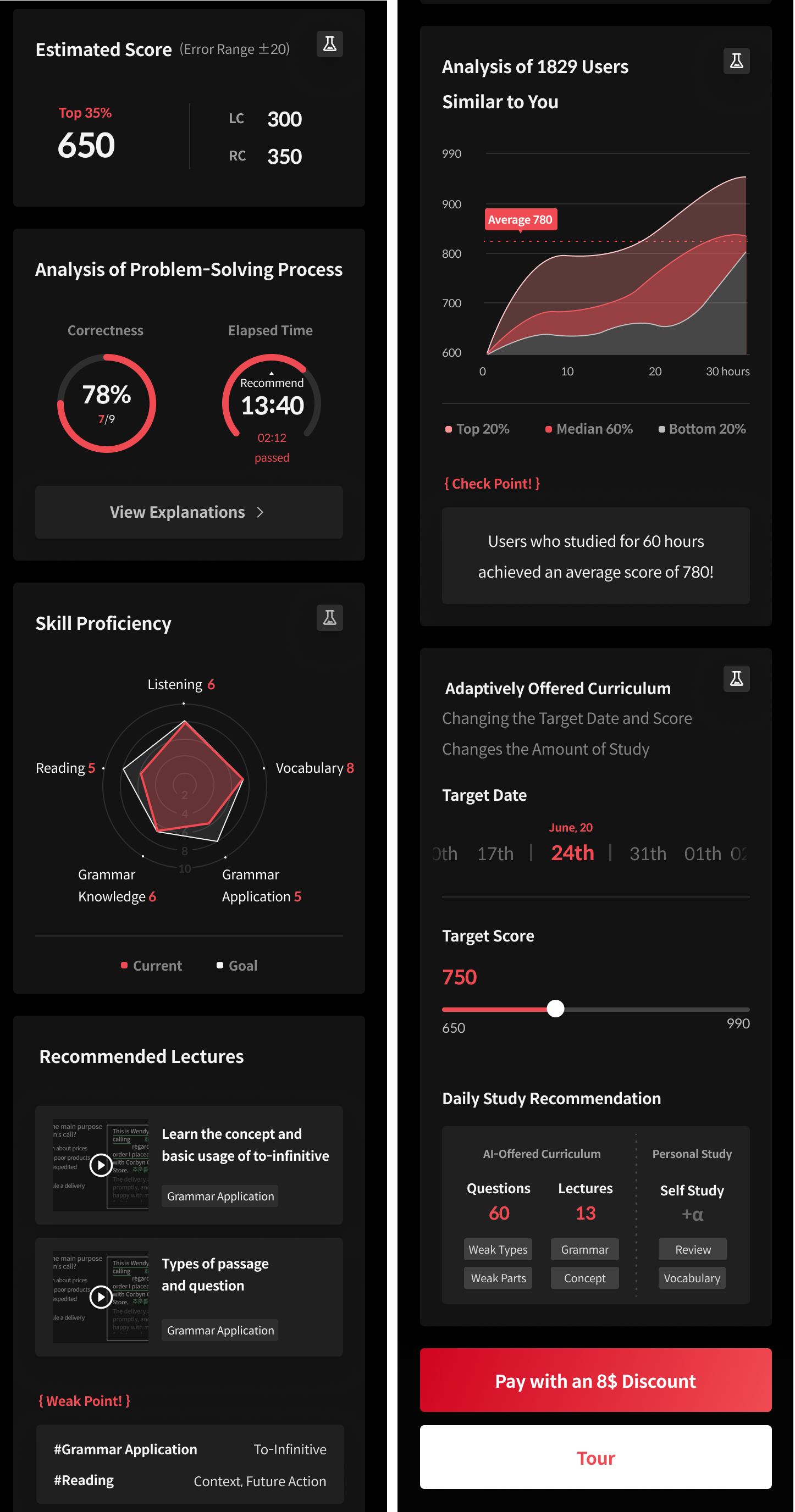}
  \caption{}
  \label{fig:page_B}
  \end{subfigure}
  \caption{Page design A(a) and B(b) proposed in the paper.
  Note that the original version of each page is in Korean.
  We present the English version of each page design in order to facilitate the understanding of readers around the world.}
  \label{fig:page_AB}
\end{figure*}

As the user decides whether to continue their study through \emph{Santa} after viewing the diagnostic page, we consider the page design that encourages user engagement and motivates them to study further.
Throughout this section, we explore the design of the diagnostic page that can most effectively express the AI features brought by back-end AI models and better explains the user’s problem-solving process in the diagnostic test.
We propose two page designs summarizing the user’s diagnostic test result: page design A (Figure \ref{fig:page_A}) and page design B (Figure \ref{fig:page_B}).
Each page design provides analytics of the diagnostic test result at different levels of information and explainability, and is powered by different AI models running behind \emph{Santa}.
The effectiveness of each page design and its impact on user engagement is investigated through controlled A/B tests in Section \ref{sec:exp}.

\subsection{Page Design A}
Page design A presents the following four components: \emph{Estimated Score}, \emph{Grade by Part}, \emph{Comparison to Users in the Target Score Zone} and \emph{Tutor’s Comment}.

\begin{figure*}
  \centering
  \begin{subfigure}[ht]{0.33\columnwidth}
  \centering
  \includegraphics[width=1\columnwidth]{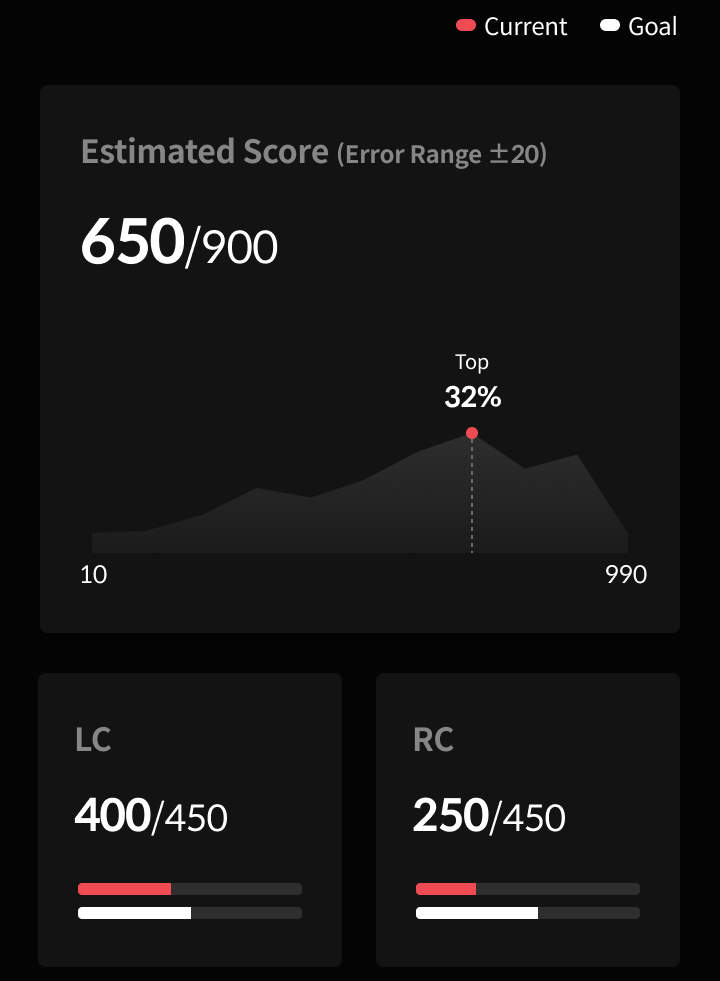}
  \caption{}
  \label{fig:page_A_score}
  \end{subfigure}
  \begin{subfigure}[ht]{0.33\columnwidth}
  \centering
  \includegraphics[width=1\columnwidth]{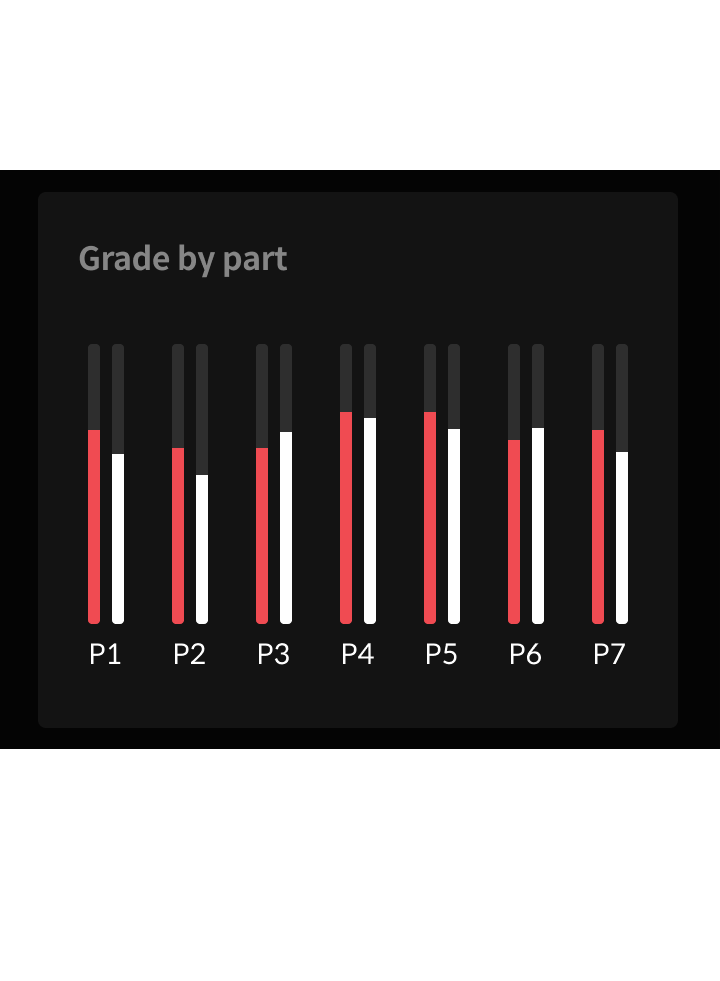}
  \caption{}
  \label{fig:page_A_part}
  \end{subfigure}
  \begin{subfigure}[ht]{0.33\columnwidth}
  \centering
  \includegraphics[width=1\columnwidth]{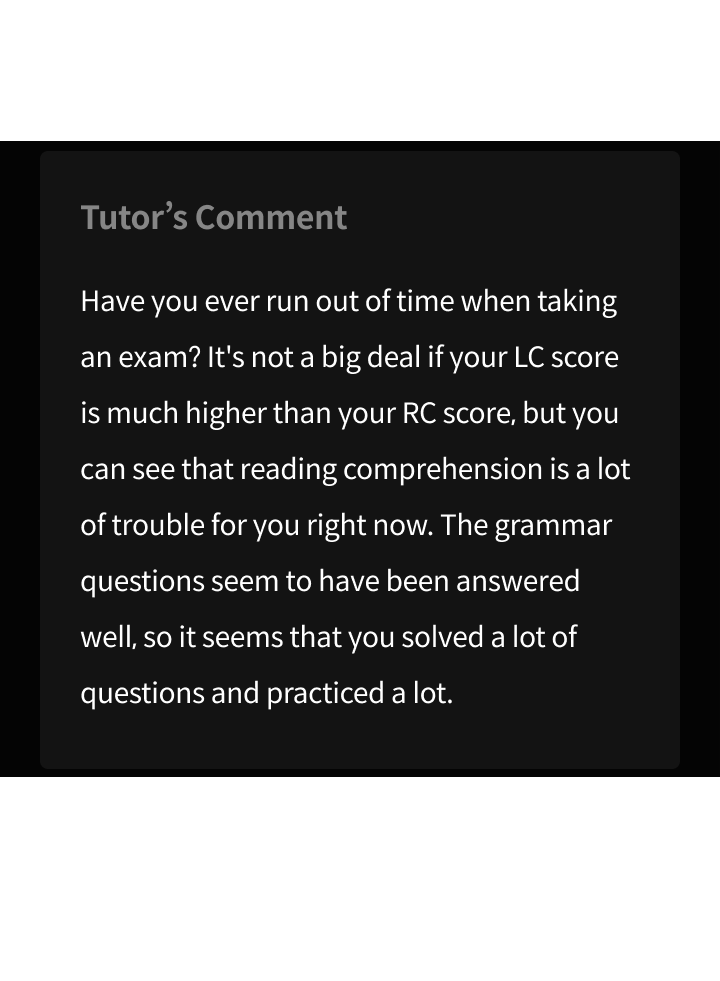}
  \caption{}
  \label{fig:page_A_tutor_comment}
  \end{subfigure}
  \caption{\emph{Estimated Score}, \emph{Grade by Part} and \emph{Tutor's Comment} components in the page design A.}
\end{figure*}

\subsubsection{Estimated Score}
This component shows the user’s overall expected performance on the actual TOEIC exam and presents the following features: estimated scores, target scores and percentile rank (Figure \ref{fig:page_A_score}).
The estimated scores are computed from the back-end score estimation model and the target scores are values the user entered before the diagnostic test.
The estimated scores and the target scores are presented together so that the user easily compares them.
The percentile rank is obtained by comparing the estimated score with scores of more than a million users recorded in the database of \emph{Santa}.

\subsubsection{Grade by Part}
This component provides detailed feedback for the user’s ability on each question type to help them identify their strengths and weaknesses (Figure \ref{fig:page_A_part}).
For each part in TOEIC exam, the red and white bar graphs show the user’s current proficiency level and the required proficiency level to achieve the target score, respectively. 
The red bar graphs are obtained by averaging the estimated probabilities of the user correctly answering the potential questions for each part.
Similarly, the white bar graphs are obtained by computing the averaged correctness probabilities for each part of users in the target score zone.

\begin{figure*}
  \centering
  \includegraphics[width=0.4\columnwidth]{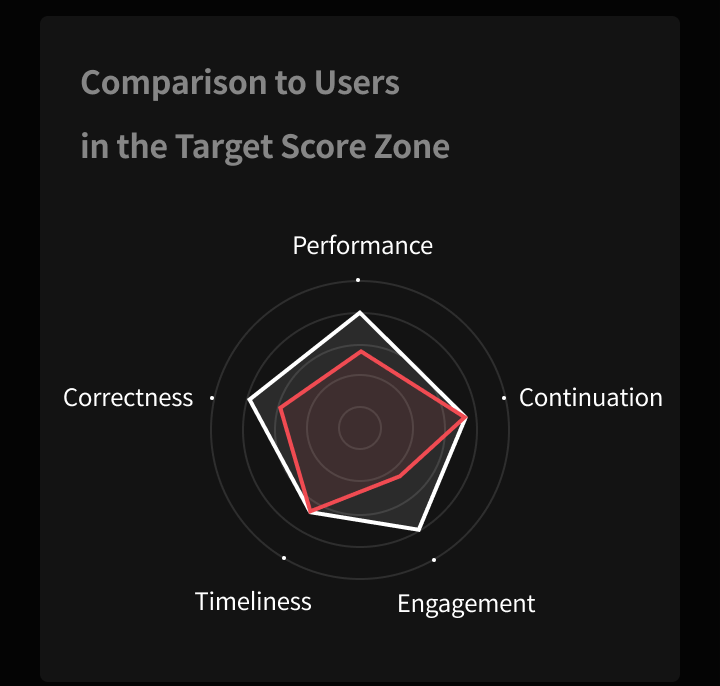}
  \caption{\emph{Comparison to Users in the Target Score Zone} in the page design A.
  The radar chart is intended to give the feeling that the AI teacher is analyzing the users closely from multiple perspectives by presenting the five features of their particular aspects of ability.}
  \label{fig:page_A_radar_chart}
\end{figure*}

\subsubsection{Comparison to Users in the Target Score Zone}
This component shows a radar chart of five features representing the user’s particular aspects of ability (Figure \ref{fig:page_A_radar_chart}).
The five features give explanations of how AI models analyze the user’s problem-solving process, making \emph{Santa} looks more like an AI teacher.
The five features are the followings:
\begin{itemize}
\item Performance: The user’s expected performance on the actual TOEIC exam.
\item Correctness: The probability that the user will correctly answer each given question.
\item Timeliness: The probability that the user will solve each given question under time limit.
\item Engagement: The probability that the user will continue studying with \emph{Santa}.
\item Continuation: The probability that the user will continue the current learning session.
\end{itemize}
The red and white pentagons present the five features with values of the current user and averaged values of users in the target score zone, respectively.
This component is particularly important as shown in Section \ref{sec:exp} that users’ engagement factors vary greatly depending on the presence or absence of the radar char.

\subsubsection{Tutor's Comment}
This component presents natural language text describing the user’s current ability and suggestions for achieving the target score (Figure \ref{fig:page_A_tutor_comment}).
This feature is intended to provide a learning experience of being taught by a human teacher through a more human-friendly interaction.
Based on the user’s diagnostic test result, the natural language text is selected among a set of pre-defined templates.

\subsection{Page Design B}
Although the page design A is proposed to provide AI-powered feedback for the user’s diagnostic test result, it has limitations in that the composition is difficult to deliver detailed information and insufficient to contain all the features computed by AI models.
To this end, the page design A is changed in the direction of making better use of AI features, leading to the page design B which is more informative and explainable to the user’s problem-solving process in the diagnostic test.
The page design B consists of the following seven components: \emph{Estimated Score}, \emph{Analysis of Problem-Solving Process}, \emph{Skill Proficiency}, \emph{Recommended Lectures}, \emph{Analysis of Users Similar to You}, \emph{Adaptively Offered Curriculum} and \emph{Santa Labs}.

\begin{figure*}
  \centering
  \begin{subfigure}[ht]{0.33\columnwidth}
  \centering
  \includegraphics[width=1\columnwidth]{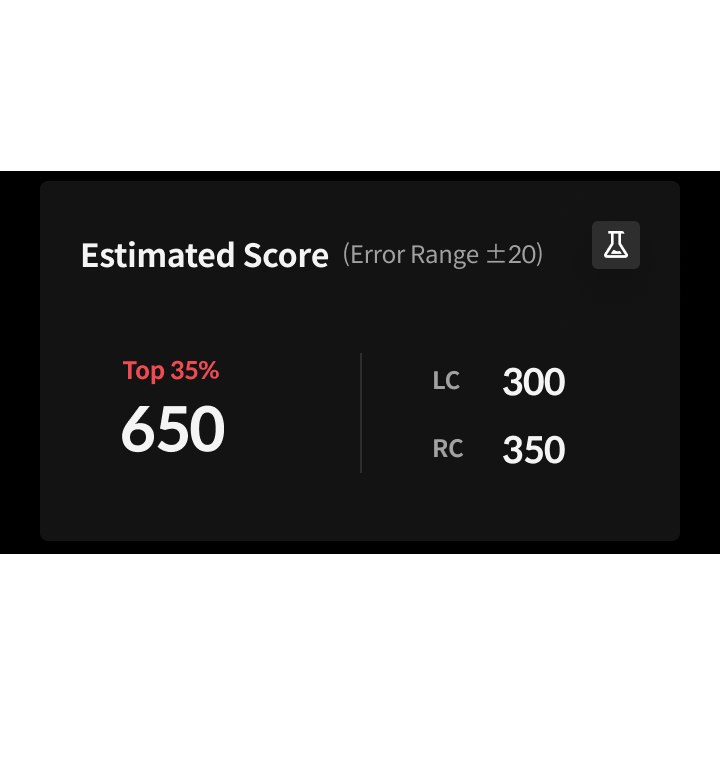}
  \caption{}
  \label{fig:page_B_score}
  \end{subfigure}
  \centering
  \begin{subfigure}[ht]{0.33\columnwidth}
  \centering
  \includegraphics[width=1\columnwidth]{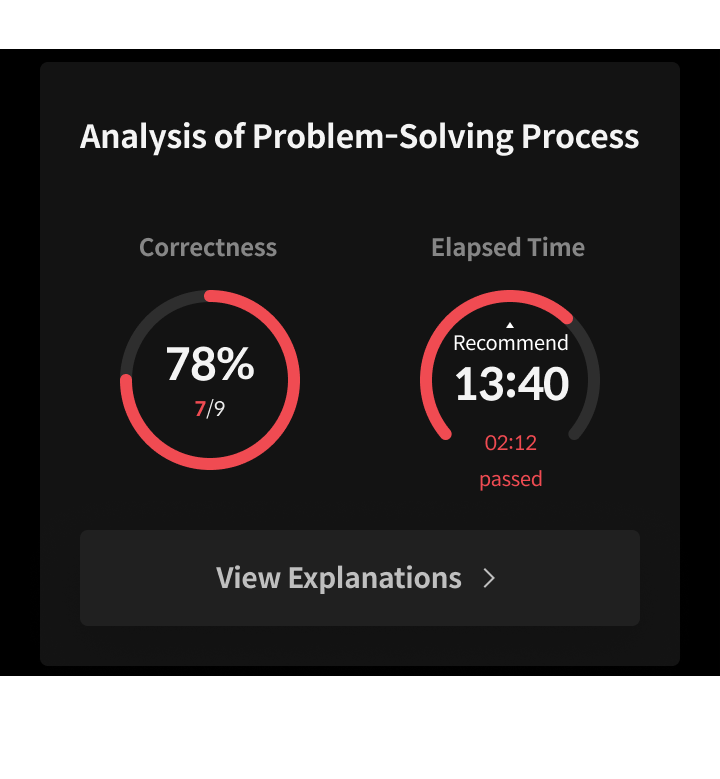}
  \caption{}
  \label{fig:page_B_problem_solving}
  \end{subfigure}
  \begin{subfigure}[ht]{0.33\columnwidth}
  \centering
  \includegraphics[width=1\columnwidth]{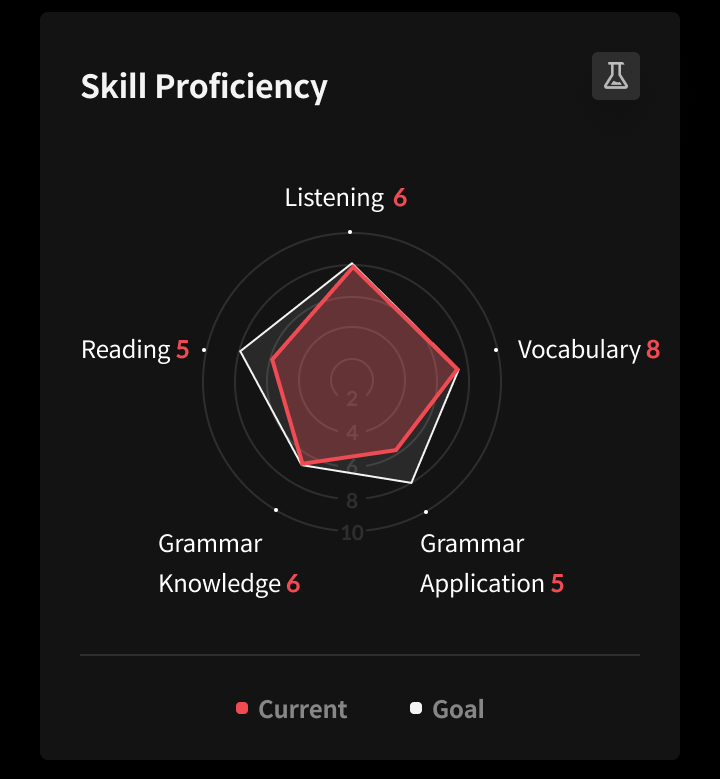}
  \caption{}
  \label{fig:page_B_proficiency}
  \end{subfigure}
  \caption{\emph{Estimated Score}, \emph{Analysis of Problem-Solving Process} and \emph{Skill Proficiency} components in the page design B.}
\end{figure*}

\subsubsection{Estimated Score}
The target scores and the mountain shaped graphic illustrating the percentile rank in the \emph{Estimated Score} component of the page design A are excluded in that of the page design B (Figure \ref{fig:page_B_score}).
The \emph{Estimated score} component of the page design B only shows the estimated scores and the number indicating the percentile rank, making this component more concise and intuitive.

\subsubsection{Analysis of Problem-Solving Process}
This component provides an overall review session for the diagnostic test (Figure \ref{fig:page_B_problem_solving}).
It presents the percentage of correct answers, the time taken to complete the diagnostic test and how much time has passed than the recommended time.
Also, through the \emph{View Explanations} button, the user can review the questions in the diagnostic test and their explanations.

\subsubsection{Skill Proficiency}
This component shows the user’s current proficiency level on each skill for TOEIC exam, making AI’s analysis of diagnostic test result more transparent and explainable (Figure \ref{fig:page_B_proficiency}).
The radar chart represents proficiency levels on a scale of 1 to 10 for the following five skills: listening, reading, grammar knowledge, grammar application and vocabulary.
Each proficiency level is obtained by normalizing the average estimated correctness probabilities of potential questions for each skill.
The red and white pentagons present the values for the five skills of the current user and averaged values of users in the target score zone, respectively.

\begin{figure*}
  \centering
  \begin{subfigure}[ht]{0.33\columnwidth}
  \centering
  \includegraphics[width=1\columnwidth]{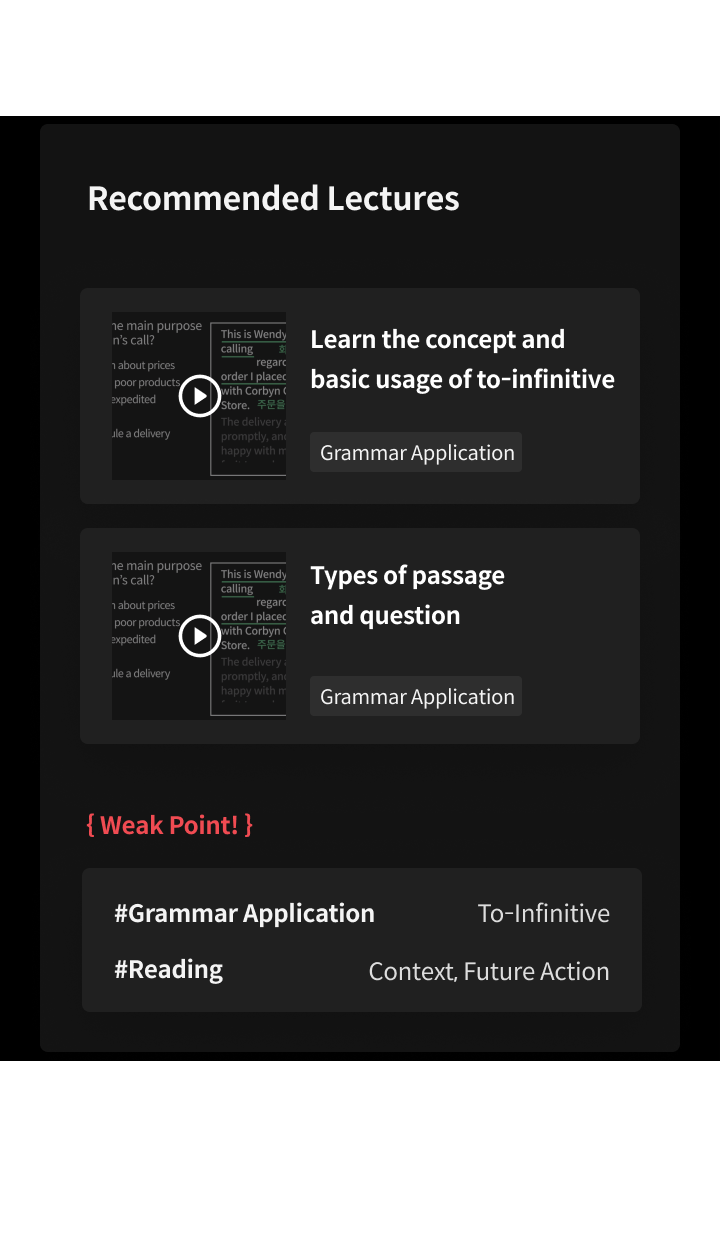}
  \caption{}
  \label{fig:page_B_lectures}
  \end{subfigure}
  \begin{subfigure}[ht]{0.33\columnwidth}
  \centering
  \includegraphics[width=1\columnwidth]{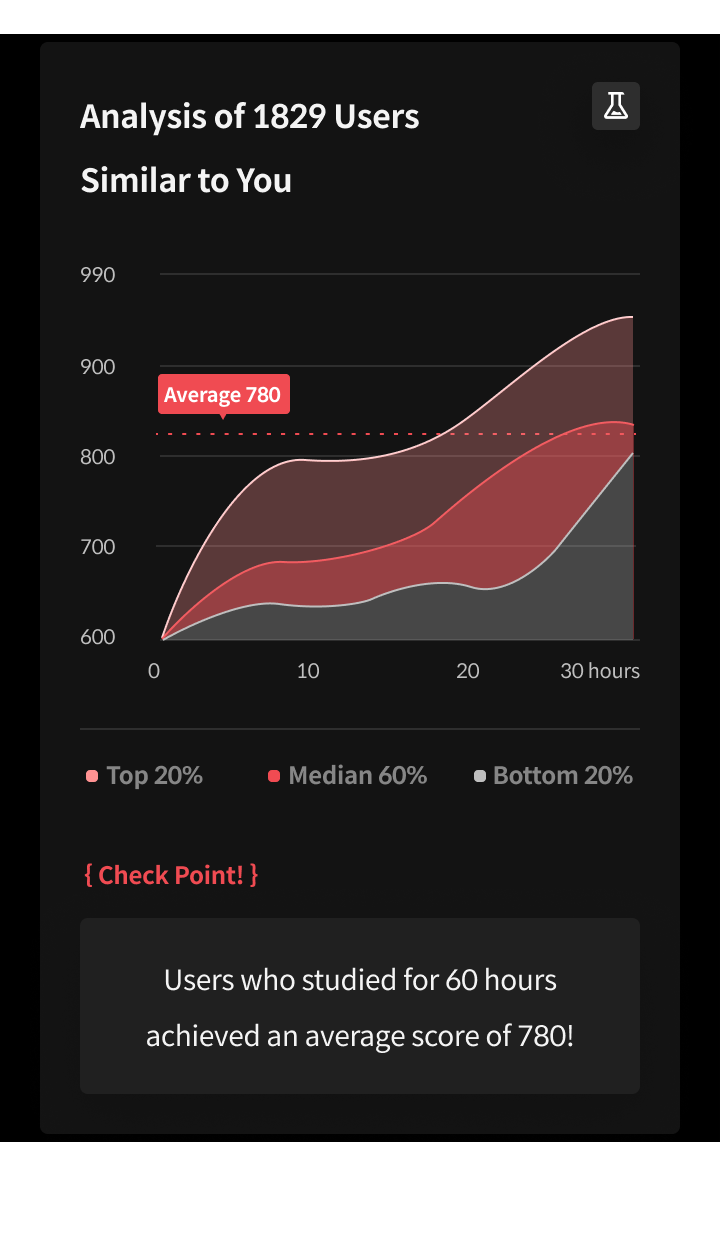}
  \caption{}
  \label{fig:page_B_analysis}
  \end{subfigure}
  \begin{subfigure}[ht]{0.33\columnwidth}
  \centering
  \includegraphics[width=1\columnwidth]{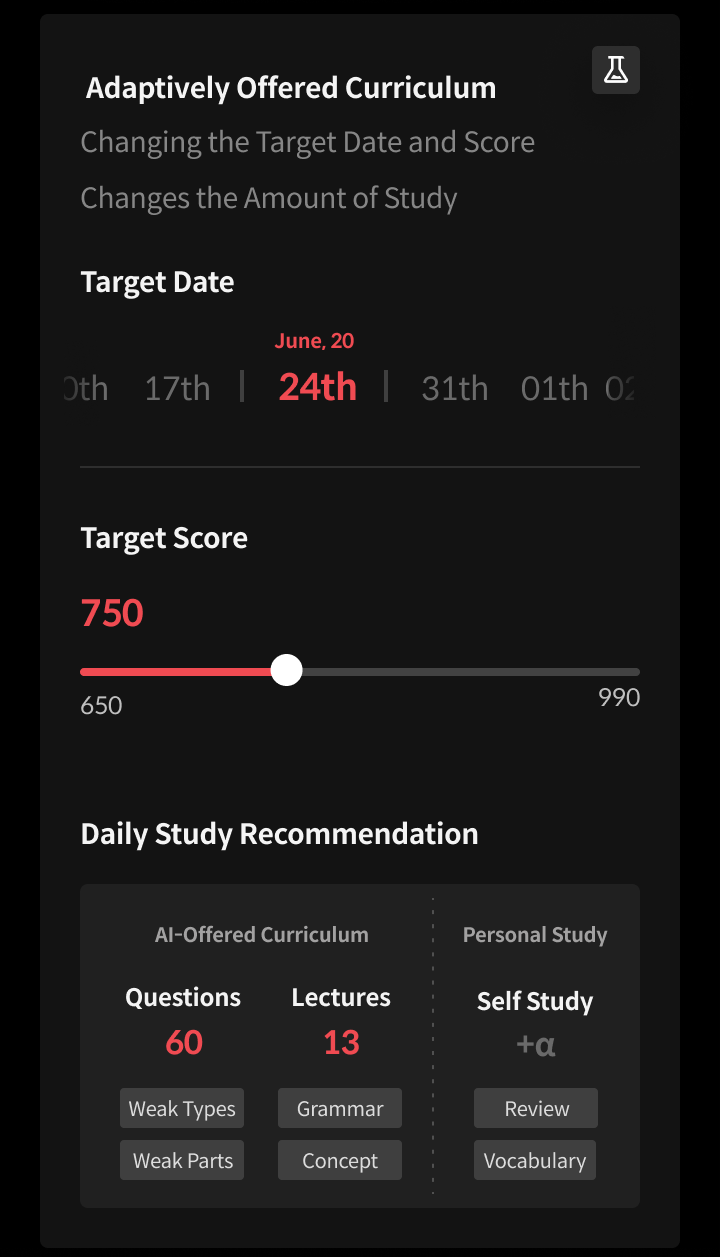}
  \caption{}
  \label{fig:page_B_curriculum}
  \end{subfigure}
  \caption{\emph{Recommended Lectures}, \emph{Analysis of Users Similar to You} and \emph{Adaptively offered curriculum} components in the page design B.}
\end{figure*}

\subsubsection{Recommended Lectures}
This component helps the user identify their weaknesses and suggests lectures to complement (Figure \ref{fig:page_B_lectures}).
Among the five skills in the \emph{Skill Proficiency} component, two skills with the lowest proficiency and their sub-skills are presented and two lectures on these skills are recommended.

\subsubsection{Analysis of Users Similar to You}
This component provides information of the change in average scores of \emph{Santa} users at the similar level to the current user, conveying the feeling that the specific score can be attained by using \emph{Santa} (Figure \ref{fig:page_B_analysis}).
It shows how the scores of the \emph{Santa} users change by dividing them into top 20\%, median 60\% and bottom 20\%, and presents the estimated average score attained after studying with \emph{Santa} for 60 hours.
This feature is obtained by finding \emph{Santa} users with the same estimated score as the current user and computing the estimated score every time they consume a learning item.

\subsubsection{Adaptively Offered Curriculum}
This component presents the learning path personalized to the user to achieve their learning objective (Figure \ref{fig:page_B_curriculum}). 
When the user changes the target date and target score by swiping, \emph{Santa} dynamically suggests the number of questions and lectures the user must study per day based on their current position.
The amount of study the user needs to consume every day is computed by finding \emph{Santa} users whose initial state is similar to the current user and tracking how their learning progresses so that the user can achieve the target score on the target date.

\begin{figure*}
  \centering
  \includegraphics[width=0.8\columnwidth]{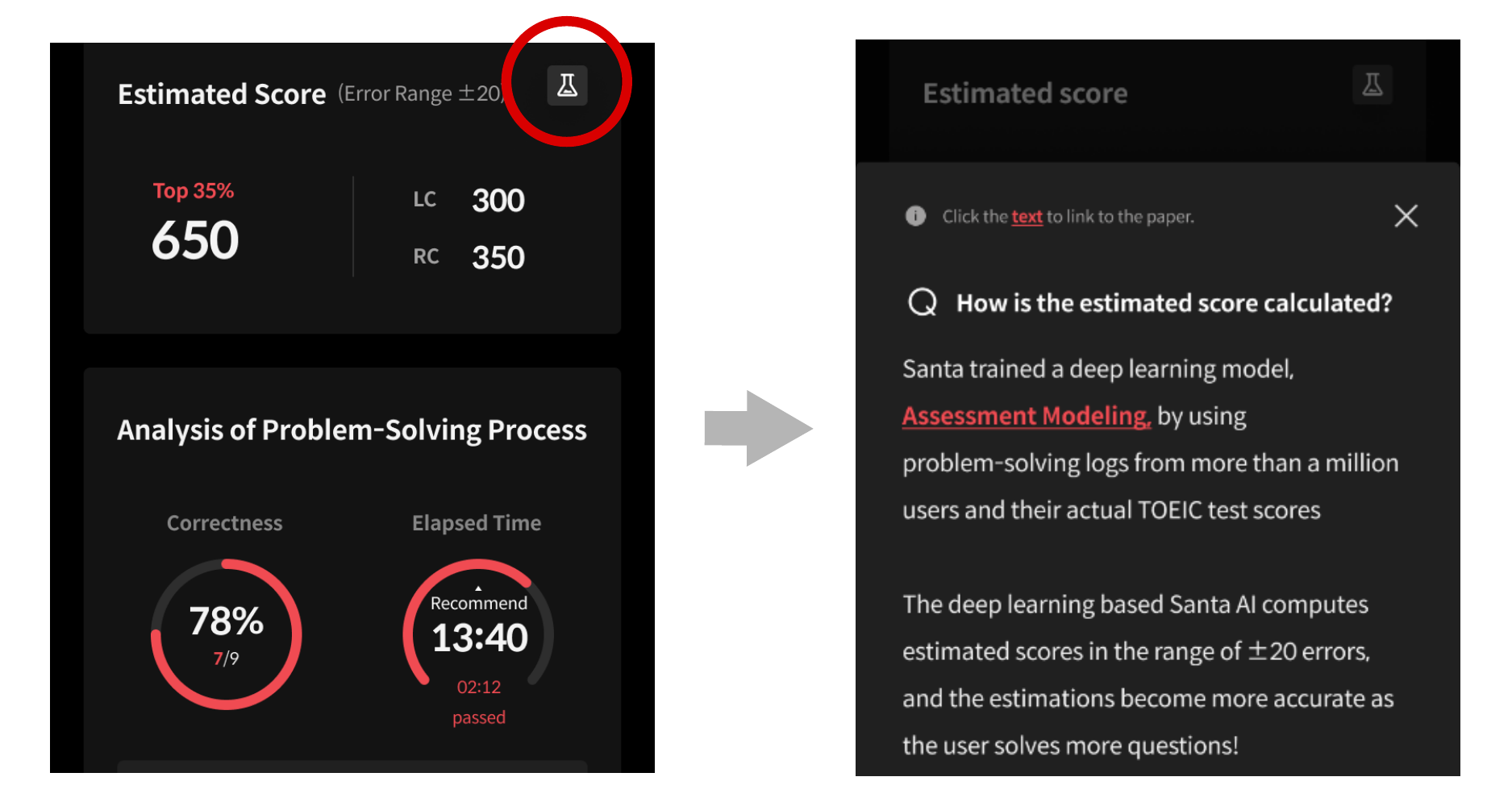}
  \caption{\emph{Santa Labs} component in the page design B.
  When the user presses the flask button next to the \emph{Estimated Score} component, a window appears with an explanation of how the estimated scores are calculated.}
  \label{fig:page_B_Santa_labs}
\end{figure*}

\subsubsection{Santa Labs}
When the user presses the flask button next to each component, a window pops up and provides an explanation of AI models used to compute the features of the component (Figure \ref{fig:page_B_Santa_labs}).
For instance, when the user presses the flask button next to the \emph{Estimated Score} component, a window appears with an explanation about the Assessment Modeling \cite{choi2020assessment}, the \emph{Santa}'s score estimation modeling method.
This component conveys information about the AI technology provided by \emph{Santa} to the user, giving them a feeling that the AI is actually analyzing them, increasing the credibility of the system.

\subsection{Back-end AI Engine}
The features in the components of each page design are computed by processing the output of \emph{Santa}’s AI engine, which takes the user's past learning activities and models individual users.
Whenever the user consumes a learning item suggested by \emph{Santa}, the AI engine updates models of individual users and makes predictions on specific aspects of their ability.
The predictions that the AI engine makes include the followings: response correctness, response timeliness, score, learning session dropout and engagement.

The response correctness prediction is made by following the approaches introduced in \cite{lee2016machine} and \cite{choi2020towards}.
\cite{lee2016machine} is the Collaborative Filtering (CF) based method which models users and questions as low-rank matrices. Each vector in the user matrix and question matrix represents latent traits of each user and latent concepts for each question, respectively.
SAINT \cite{choi2020towards} is a deep learning based model that follows the Transformer \cite{vaswani2017attention} architecture.
The deep self-attentive computations in SAINT allows to capture complex relations among exercises and responses.
Since the CF-based model can quickly compute the probabilities of response correctness for the entire questions of all users and SAINT predicts the response correctness probability for each user with high accuracy, the two models are complementary to each other in real world applications where both accuracy and efficiency are important.

Assessment Modeling (AM) \cite{choi2020assessment} is a pre-train/fine-tune approach to address the label-scarce educational problems, such as score estimation and review correctness prediction.
Following the pre-train/fine-tune method proposed in AM, a deep bidirectional Transformer encoder \cite{devlin2018bert} based score estimation model is first pre-trained to predict response correctnesses and timelinesses of users conditioned on their past and future interactions, and then fine-tuned to predict scores of each user.
The response timeliness and score are predicted from the pre-trained model and the fine-tuned model, respectively.

The learning session dropout prediction is based on the method proposed in DAS \cite{lee2020deep}.
DAS is a deep learning based dropout prediction model that follows the Transformer architecture.
With the definition of session dropout in a mobile learning environment as an inactivity for 1 hour, DAS computes the probability that the user drops out from the current learning session whenever they consume each learning item.

The engagement prediction is made by the Transformer encoder based model.
The model is trained by taking the user’s learning activity record as an input and matching the payment status based on the assumption that the user who makes the payment is engaged a lot with the system.
\section{Experimental Studies} \label{sec:exp}
In this section, we provide supporting evidence that AI-driven interface design for ITS promotes student engagement by empirically verifying the followings through the real world application: 1) the impact of the radar chart in the page design A on user engagement, and 2) comparison of the page design A and B on user engagement.
We conduct A/B tests on new incoming users of \emph{Santa}, with the users randomly assigned either group A or B.
Both groups of users take the diagnostic test, and at the end, users in different groups are shown different diagnostic test analytics pages.
Throughout the experiments, we consider the following four factors of engagement: conversion rate, Average Revenue Per User (ARPU), total profit and the average number of free questions a user consumed. 
Monetary profits are an essential factor for evaluating the users’ engagement since paying for a service means that the users are highly satisfied with the service and requires a strong determination of actively using the service.
For users without the determination to make payment, the average number of free questions a user consumed after the diagnostic test is a significant measure of engagement since it represents their motivation to continue the current learning session.

\subsection{Impact of Radar Chart in Page Design A on Student Engagement}
\begin{figure*}[h]
  \centering
  \includegraphics[width=0.7\columnwidth]{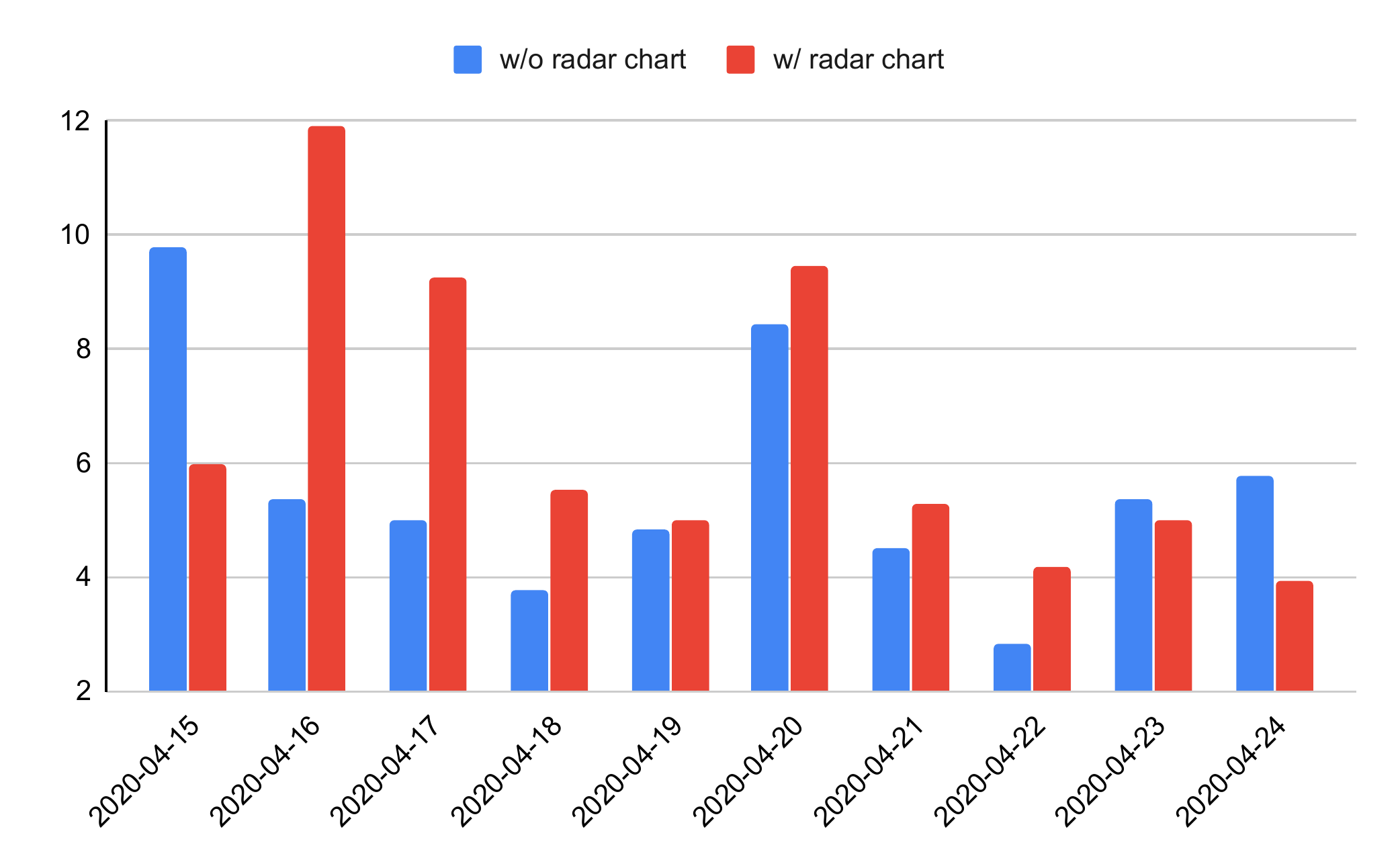}
  \caption{Comparison of the conversion rate on every day between the users of the A/B test.}
  \label{fig:daily_payment_radar_chart}
\end{figure*}

\begin{figure*}[h]
  \centering
  \includegraphics[width=0.7\columnwidth]{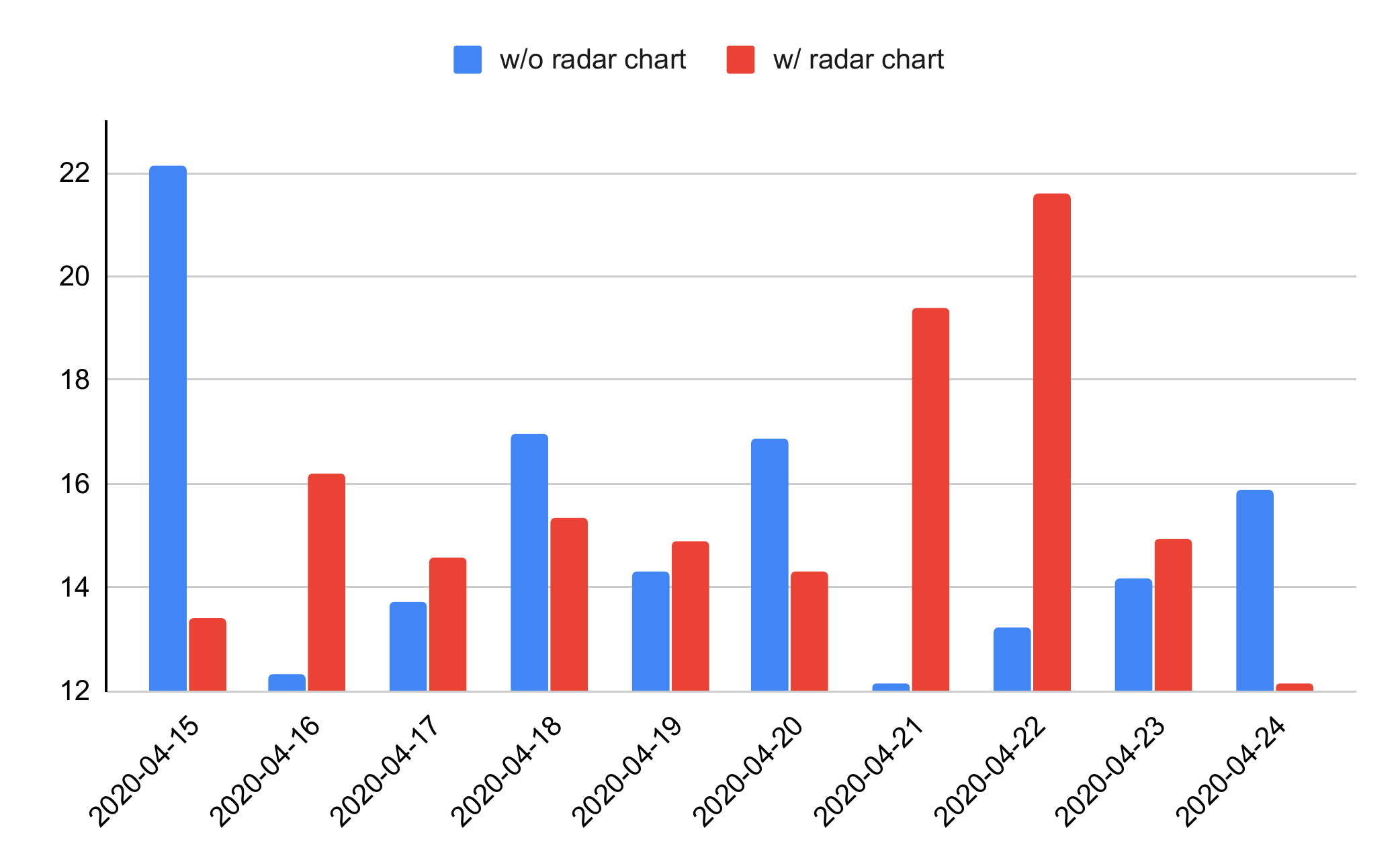}
  \caption{Comparison of the average number of free questions a user consumed on every day between the users of the A/B test.}
  \label{fig:daily_questions_radar_chart}
\end{figure*}

From April 15th to 24th, we conducted an A/B test by randomly assigning two different diagnostic test analytics pages to the users: one without the radar chart in the page design A (1,391 users) and another one the page design A (1,486 users).
Table \ref{tab:radar_chart} shows the overall results.
We see that the page design A with the radar chart improves all factors of user engagement.
With the radar chart, the conversion rate, ARPU, total profit and the average number of free questions a user consumed are increased by 22.68\%, 17.23\%, 25.13\% and 11.78\% respectively, concluding that a more AI-like interface design for ITS encourages student engagement.
Figure \ref{fig:daily_payment_radar_chart} and Figure \ref{fig:daily_questions_radar_chart} show the comparison of the conversion rate and the average number of free questions a user consumed per day between the users of the A/B test, respectively.
We observe that the users of the page design A with the radar chart made more payments and solved more free questions throughout the A/B test time period.

\begin{table}[h]
\caption{Impacts of the radar chart on user engagement.}
\centering
\begin{tabular}{ccc}
\toprule
& w/o radar chart & w/ radar chart \\
\hline
Conversion rate (\%) & 5.25 & 6.26 \\
ARPU (\$) & 5.92 & 6.94 \\
Total profit (\$) & 8,236.01 & 10,305.58 \\
\# of free questions consumed & 14.77 & 16.51 \\
\bottomrule
\end{tabular}
\label{tab:radar_chart}
\end{table}

\subsection{Comparison of Page Design A and B on user engagement}
\begin{figure*}[h]
  \centering
  \includegraphics[width=0.7\columnwidth]{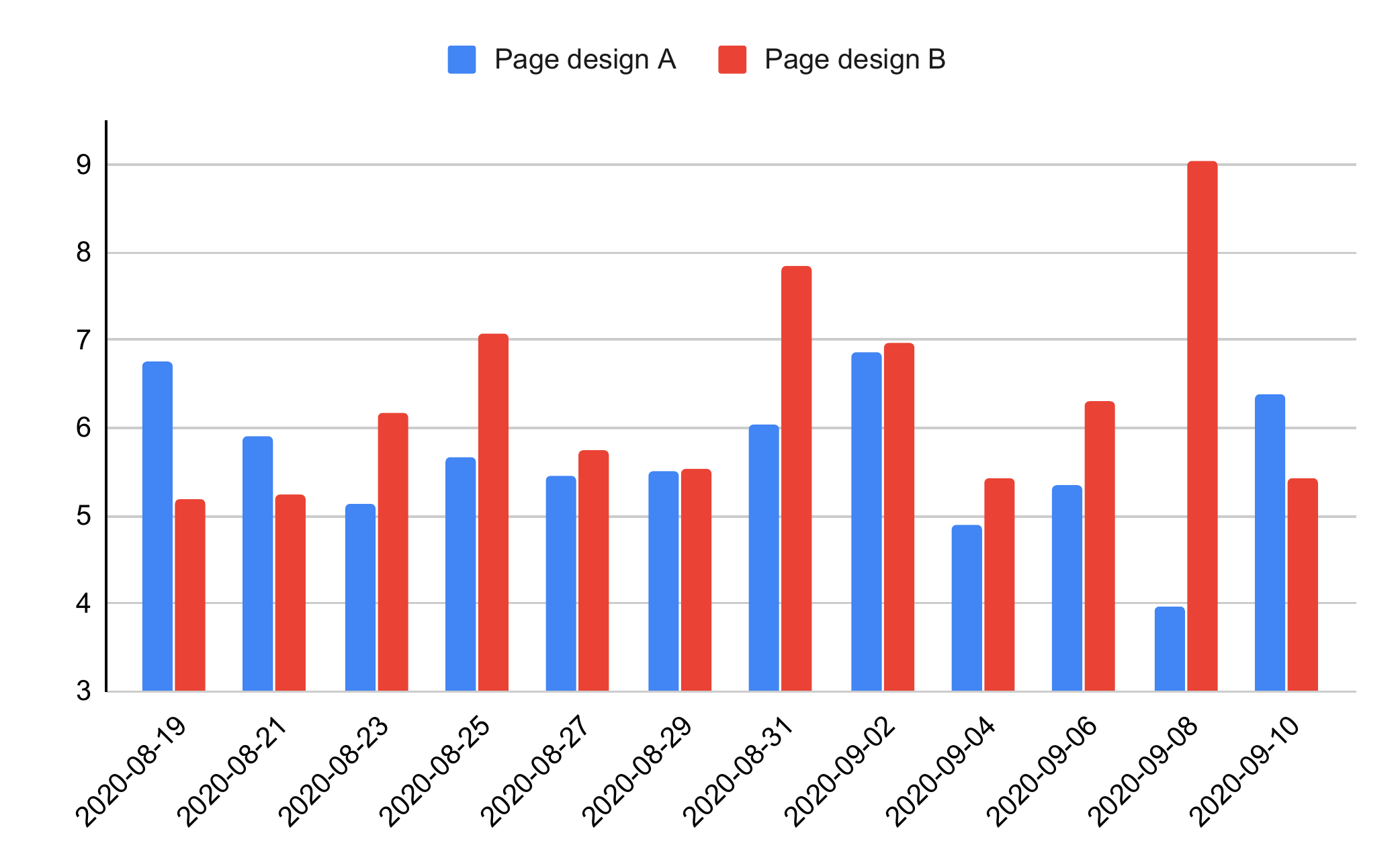}
  \caption{Comparison of the conversion rate on every two days between the users of the A/B test.}
  \label{fig:daily_payment_design_AB}
\end{figure*}

\begin{figure*}[h]
  \centering
  \includegraphics[width=0.7\columnwidth]{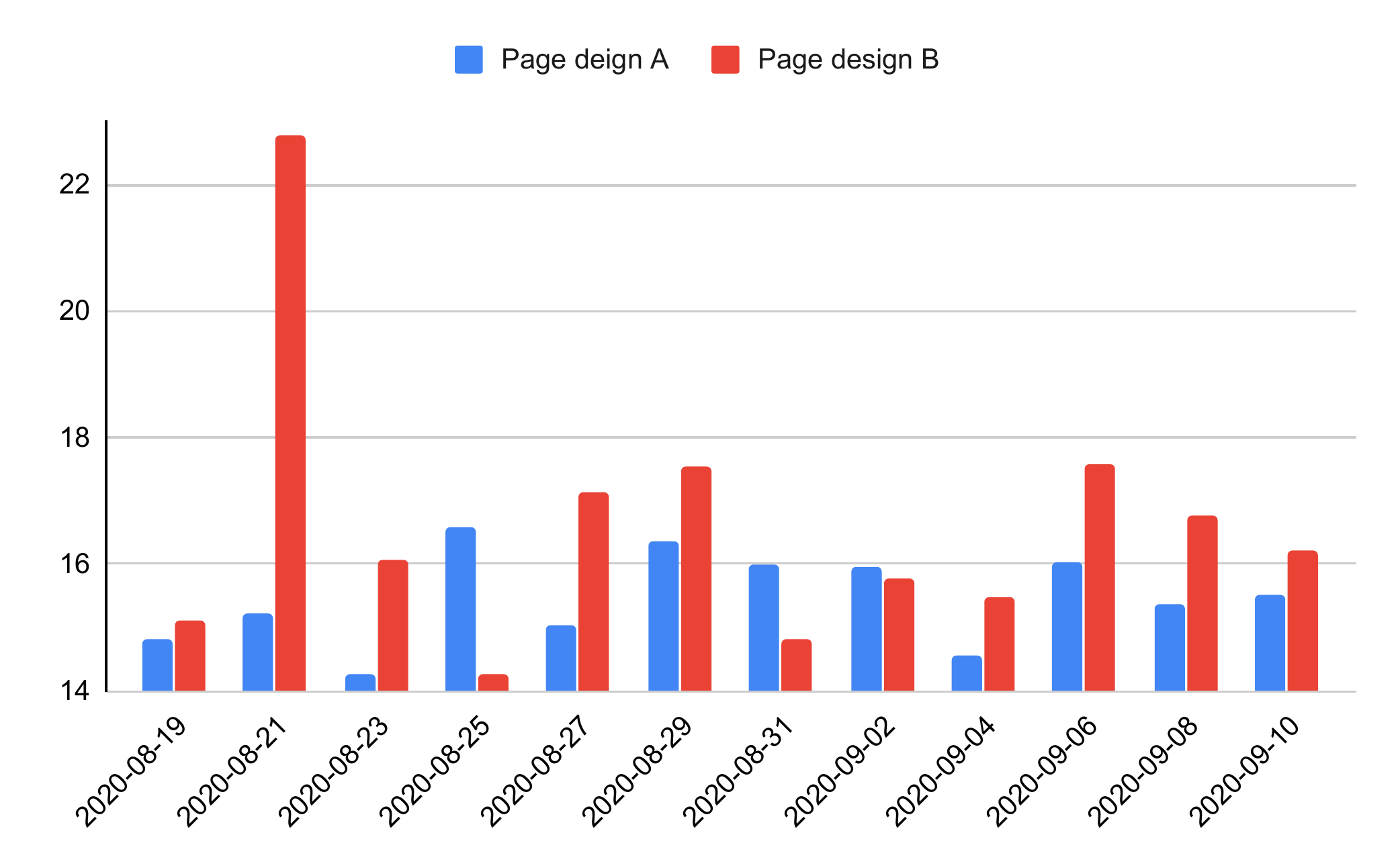}
  \caption{Comparison of the average number of free questions a user consumed on every two days between the users of the A/B test.}
  \label{fig:daily_questions_design_AB}
\end{figure*}

The A/B test of the page design A and B was conducted from August 19th to September 11th by randomly allocating them to the users.
9,442 users were allocated to the page design A and 9,722 users were provided the page design B.
The overall results are shown in Table \ref{tab:AB}.
Compared to the page design A, the page design B is better at promoting all factors of user engagement by increasing 11.07\%, 10.29\%, 12.57\% and 7.19\% of the conversion rate, ARPU, total profit and the average number of free questions a user consumed, respectively.
Note that although the page design with the radar chart in the previous subsection and the page design A are the same, there is a difference between the values of the engagement factors of page design with the radar chart in Table \ref{tab:radar_chart}, and those of the page design A in Table \ref{tab:AB}.
The absolute value of each number can be changed by external factors, such as timing and the company's public relations strategy, and these external factors are not a problem as they apply to both A and B groups in the A/B test.
The comparisons of the conversion rate and the average number of free questions a user consumed on every two days between the users assigned to the page design A and B are presented in Figure \ref{fig:daily_payment_design_AB} and Figure \ref{fig:daily_questions_design_AB}, respectively.
We can observe in the figures that users experiencing the page design B made more payments and solved more free questions during the A/B test time period.
Throughout the experiment, the results show that a more informative and explainable design of interface for ITS by making better use of AI features improves student engagement.

\begin{table}[h]
\caption{Impacts of the page design A and B on user engagement.}
\centering
\begin{tabular}{ccc}
\toprule
& Page design A & Page design B \\
\hline
Conversion rate (\%) & 5.60 & 6.22 \\
ARPU (\$) & 5.54 & 6.11 \\
Total profit (\$) & 83,335.76 & 93,807.27 \\
\# of free questions consumed & 15.15 & 16.24 \\
\bottomrule
\end{tabular}
\label{tab:AB}
\end{table}

\section{Conclusions and Future Work}
We have investigated the effects of AI-driven interface design for ITS.
In particular, we hypothesized that diagnostic page design summarizing analytics for students' problem-solving process that makes better use of AI features would encourage student engagement.
For this, we proposed several page designs that couple the interface with AI features in different levels and empirically verified their impacts on student engagement.
We have conducted A/B tests on new students using an active mobile ITS \emph{Santa}.
We considered conversion rate, Average Revenue Per User (ARPU), total profit and the average number of free questions a user consumed as factors measuring the degree of engagement.
The A/B test results showed that the page designs that effectively expresses the AI features brought by back-end AI engine and thus better explain analysis of the user’s diagnostic test result promote all factors of engagement, concluding that AI-driven interface design improves student engagement.
Avenues of future work include 1) updating a page summarizing the AI’s analysis on the user in the learning session after the diagnostic test. 2) finding more interface designs that can further enhance student engagement by making good use of, expressing and utilizing AI features.

\section{Related Works}
\begin{comment}
The work of this paper is related to the following two categories of research areas: UI for ITS and and explainability of AIEd.

Explaining what exactly makes AI models arrive at their predictions and making them transparent to users is an important issue \cite{gunning2017explainable,gunning2019darpa,dove2020monsters}, and have been actively studied in both HCI \cite{abdul2018trends,kizilcec2016much,stumpf2009interacting,wang2019designing} and ML \cite{samek2017explainable} community.
There are lots of works about the issue of explainability in many subfields of AI including computer vision \cite{norcliffe2018learning,fong2017interpretable,kim2017interpretable,zhang2018interpretable}, natural language processing \cite{lei2017interpretable,fyshe2015compositional,jiang2018interpretable,panigrahi2019word2sense} and speech processing \cite{ravanelli2018interpretable,korzekwa2019interpretable,sun2020fully,tan2015improving}.
In this section, we narrow down the scope and describe how the AIEd researchers have addressed the issue.
\end{comment}

\subsection{Design of UI for ITS}
Although the development of ITS has become an active area of research in recent years, most of the studies have mainly focused on learning science, cognitive psychology and artificial intelligence, resulting in little works done in the context of UI.
\cite{granic2000user} describes the UI issue of an intelligent authoring shell, which is an ITS generator.
Through experiences in the usage of TEx-Sys, an authoring shell proposed in the paper, the authors discusses the importance of a well designed UI that brings system functionality to users.
\cite{glavinic2001interacting} considers applying multiple views to UI for ITSs.
The paper substantiates the usage of multiple perspectives on the domain knowledge structure of an ITS by the implementation of MUVIES, a multiple views UI for ITS.
Understanding students' emotional states has become increasingly important for motivating their learning.
Several works \cite{lin2014influence,lin2014usability} incorporate affective interface in ITS to monitor and correct the students' states of emotion during learning.
\cite{lin2014influence} studies the usage affective ITS in Accounting remedial instruction.
Virtual agents in the system analyze, behave and give feedback appropriately to students' emotions to motivate their learning.
\cite{lin2014usability} proposes ATSDAs, an affective ITS for digital arts.
ATSDAs analyzes textual input of a student to identify emotion and learning status of them.
A visual agent in the system adapts to the student, provides text feedback based on the inferred results and thereby increases their learning interest and motivation.
The performance of a software can be measured by its usability, a quality that quantifies ease of use. Whether applying usability testing and usability principles to the design of UI can improve the performance of ITS is an open question \cite{chughtai2015usability}.
\cite{koscianski2014design} discusses the importance of UI design, usability and software requirements and suggests employing heuristics from software engineering and learning science domains in the development process of ITS.
An example of applying basic usability techniques to the development and testing of ITS is presented in \cite{roscoe2014writing}.
The paper introduces Writing Pal, an ITS for helping to improve students’ writing proficiency.
The design of Writing Pal includes many usability engineering methods, such as internal
usability testing, focus groups and usability experiments.

\subsection{Explainability in AIEd}
Providing an explainable feedback which can identify strengths and weaknesses of a student is a fundamental task in many educational applications \cite{conati2018ai}.
DIRT \cite{cheng2019dirt} and NeuralCDM \cite{wang2019neural} propose methods to enhance explainability of educational systems through a cognitive diagnosis modeling, which aims to discover student’s proficiency levels on specific knowledge concepts.
DIRT incorporates neural networks to compute parameters of Item Response Theory (IRT) model. With the great feature representation learning power of neural networks, DIRT could learn complex relations between students and exercises, and give explainable diagnosis results.
A similar approach is taken in NeuralCDM.
However, the diagnosis model of NeuralCDM is an extended version of IRT with monotonicity assumption imposed on consecutive fully-connected neural network layers before the final output.
Deep-IRT \cite{yeung2019deep} is a synthesis of IRT and DKVMN \cite{zhang2017dynamic}, a memory-augmented neural networks based knowledge tracing model.
Deep-IRT leverages intermediate computations of DKVMN to estimate the item difficulty level and the student ability parameters of IRT model.
EKT, proposed in \cite{huang2019ekt}, is a bidirectional LSTM based knowledge tracing model.
EKT explains the change of knowledge mastery levels of a student by modeling evolution of their knowledge state on multiple concepts over time.
Also, equipped with the attention mechanism, EKT quantifies the relative importance of each exercise for the mastery of the student’s multiple knowledge concepts.

As pointed in \cite{manouselis2012recommender}, explainability also poses challenges to educational recommender systems.
\cite{barria2019explaining} addresses this issue by providing a visual explanation interface composed of concepts’ mastery bar chart, recommendation gauge and textual explanation.
When a certain learning item is recommended, the concepts’ mastery bar chart shows concept-level knowledge of a student, the recommendation gauge represents suitability of the item and the textual explanation describes the recommendation rule why the item is suggested.
Rocket, a tinder-like UI introduced in \cite{choi2020choose}, also provides explainability in learning contents recommendation.
When an ITS proposes a learning material to a user, Rocket shows a polygonal visual summary of AI-extracted features, such as the probability of the user correctly answering the question being presented and expected score gain when the user correctly answers the question, which gives the user insight into why the system recommends the learning material.
Based on the AI-extracted features, the user can decide whether to consume the suggested learning material or not through swiping or tapping action.

\bibliographystyle{ACM-Reference-Format}
\bibliography{ref}

\end{document}